\documentclass[a4paper, 11pt]{article}
\usepackage[utf8]{inputenc}
\usepackage{jcappub}
\usepackage{amsfonts}
\usepackage{graphicx}
\usepackage{amssymb}
\usepackage{amsmath}
\usepackage{breqn}
\usepackage{array}
\usepackage{float}
\usepackage{tikz}
\usepackage{subfig}
\usetikzlibrary{shapes.geometric, arrows}
\tikzstyle{st} = [rectangle, rounded corners, text width = 3cm, text centered, draw = black ]
\tikzstyle{arrow} = [->,>=stealth]

\title{Stable contraction in Brans-Dicke cosmology}

\author{Debottam Nandi}
\affiliation{Department of Physics, Indian Institute of
  Technology Madras, Chennai 600036, India}

\emailAdd{debottam@physics.iitm.ac.in}

\abstract{
Contracting Universe (including bouncing models) solution generally depends on the initial conditions and hence possesses an extreme fine-tuning problem. In order to probe the stability of those solutions, in this work, we consider the Brans-Dicke theory with the power law potential in the presence of an additional barotropic matter in the homogeneous and isotropic background. We study the phase space and obtain critical points. We find that the quadratic potential is a special case where one of the solutions always gives rise to de-Sitter solution in all conformally connected frame. We generalize the condition for arbitrary power law potential and find that contracting Universe solution can indeed lead to an attractor solution. In doing so, we also provide an example of a matter contracting Universe that leads to near scale-invariant spectra and study the behavior near the fixed point. In the vicinity of this point, the system behaves as the Universe contains three different types of matter. Amongst them, with time, energy densities of two effective fluids decay down and only the leading order solution survives. Therefore, the non-minimal coupling can address and solve the problem of fine-tuning of the contraction models and may open a different outlook.
}

\begin{document}

\maketitle

\section{Introduction}

In solving non-linear differential equations, initial conditions play a crucial role in determining the dynamical behavior of the system as different initial conditions may lead to different trajectories which may behave completely differently. Therefore, analyzing the trajectories corresponding to points in a nearby vicinity is of great importance as it determines the general behavior of the solution around a region, i.e.,  the solutions with nearby initial conditions behave the same ---  {\it an attractor solution}, or the behavior drastically changes with the slight change of initial conditions --- {\it a non-attractor solution}. 

Gravity acts similarly. It is a highly non-linear system and only specific choice of initial conditions leads to a solvable system. In cosmology, for instance, by imposing homogeneous and isotropic conditions simplifies the equations drastically and by choosing appropriate model parameters, we can obtain our desired solution that is consistent with the observation. However, early Universe possesses the problem of arbitrariness of the initial condition as there is no such reason for choosing precise initial conditions. This brings a complexity in the system as different initial conditions may completely change the behavior of the desired solution required. Depending on the model, the desired solution can either be an attractor or a non-attractor if we impose arbitrary initial conditions. Even if the solution is an attractor, only the leading order solution coincides with the desired solution and other terms, too, can play a crucial role in determining the dynamics. Thus, in cosmology, the theory with the desired solution can be thought of as an effective theory where only leading order solution is considered.

In order to understand as well as to simplify the equations in cosmology, theories with power law scale factor are at most importance. In the Einstein frame, i.e. when the matter is minimally coupled to the curvature, canonical scalar field with exponential potential leads to a power law scale factor $a(t) \sim t^\beta$ in the absence of any other additional matter and the model parameter $\beta$ can determine whether power law scale factor solution is an attractor or not. It is well known that for an expanding Universe, $\beta > 1/3$ solution leads to an attractor solution whereas for contracting Universe, $\beta$ has to maintain the value of $< 1/3$ to maintain its attractor nature. For example, the dust matter solution, i.e., $\beta = 2/3$ for canonical scalar field with exponential potential are non-attractors if the Universe is contracting. However, the same solution in an expanding Universe acts as an attractor. 

Within the standard model of cosmology, the inflationary paradigm \cite{Mukhanov:1990me,Bassett:2005xm,Sriramkumar:2009kg,Baumann:2009ds,Linde:2014nna,Martin:2015dha} solves the horizon and flatness problem as well as fits the best with observations. However, perhaps the greatest achievement of the inflationary paradigm is that, most models in this paradigm solve the initial value problem, i.e., the solutions are attractors and independent of the initial conditions. Therefore, the effective theory of assuming the scale factor solution justifies as the solution can be thought of as the solution in the asymptotic limit. However, even with tighter constraints, we are unable to rule out sufficient number of models within the inflationary paradigm \cite{Martin:2010hh,Martin:2013tda,Martin:2013nzq,Martin:2014rqa,Gubitosi:2015pba}. Also, inflation being insensitive to initial conditions is still debatable. For instance,  in Ref. \cite{East:2015ggf}, using non-perturbative simlulations, authors showed that the inflationary expansion starts under very specific circumstances. In Ref. \cite{Clough:2016ymm}, authors pointed out the fact that small field potential fails to start the inflationary expansion under most initial conditions. Therefore, there is a growing interest in finding new alternatives to inflation. Amongst them, classical bouncing models are the most popular one \cite{Wands:1998yp,Novello:2008ra,Cai:2014bea,Battefeld:2014uga,Lilley:2015ksa,Ijjas:2015hcc,Brandenberger:2016vhg}. However, the problem with the most bouncing models is that, either the models are extremely difficult to construct in terms of a single scalar field (e.g., ekpyrotic model as the single scalar field action does not lead to scale-invariant spectra) or they behave as a non-attractor (e.g., matter bounce).

In this work, in order to investigate the attractor nature of the alternatives to inflation, mainly the bouncing models, we concentrate on contraction of the Universe as in general it mainly determines the fine-tuning of the system. Since, in the minimal Einstein frame, contraction always leads to a non-attractor solution\footnote{Only known attractors are the ekpyrotic models \cite{Levy:2015awa}. In fact, in Ref. \cite{Garfinkle:2008ei}, authors showed that the ekpyrotic contraction is a `super-smoother', i.e., it is robust to a very wide range of initial conditions and avoid Kasner/mixmaster chaos. A similar statement could never be proven
about any other primordial scenario.}, in this work, we concentrate on the non-minimally coupled models \cite{Horndeski1974, Deffayet2009, Nojiri:2010wj, Nojiri:2017ncd}. The simplest of them is the Brans-Dicke theory \cite{Brans-Dicke1961} where the curvature scalar is non-minimally coupled to the canonical scalar field\footnote{The frame, in general, is called the Jordan frame where curvature scalar is coupled to a generalized scalar field \cite{Faraoni:1999hp}.}. The aim of this work is two-fold: first, to study the general attractor behavior in Brans-Dicke frame, and second, to find out whether a contraction solution can act as an attractor that may also lead to (near) scale-invariant spectra \cite{1994ApJ...420..445F, Ade:2015xua, Akrami:2018odb, Aghanim:2018eyx}.

This work can be thought of as an extension of the works studied earlier \cite{Kolitch:1994kr,Kolitch:1994qa,Santos:1996jc,Holden:1998qg,Hrycyna:2013hla,Hrycyna:2013yia, Kofinas:2016fcp}. In the recent works \cite{Hrycyna:2013hla,Hrycyna:2013yia}, authors consider quadratic potential and using the two model parameters $(\omega_{\rm BD}, w_m)$, i.e., the Brans-Dicke parameter and the equation of state of the barotropic fluid, respectively, they study the general phase space structure. In doing so, the authors consider different values of $\omega_{\rm BD}$. One obvious special choice is $\omega_{\rm BD} = 0$. This value corresponds to the so called ``O'Hanlon theory" \cite{0022-3689-5-6-005} or ``massive dilaton gravity" \cite{Capozziello:2010zz} or the $f(R)$ Gravity \cite{DeFelice:2010aj}. $f(R)$ Gravity with such quadratic potential leads to the well known Starobinsky model with de-Sitter solution \cite{STAROBINSKY1982175, Starobinsky:1987zz}. Amongst many values of $\omega_{\rm BD}$, authors avoided a specific value of $\omega_{\rm BD} = -3/2$ as it leads to pathologies \cite{Dabrowski:2005yn}.

In this work, we concentrate on studying the Brans-Dicke theory in a more generalized manner. However, the objectives are different. As we have mentioned before, the motivation of this work is to study the attractor behavior of the contracting Universe (contraction phase of the bouncing Universe) models that lead to scale invariant spectra. In order to achieve that, we consider power law potential which leads to the power law scale factor in Brans-Dicke theory\footnote{In the Einstein frame, exponential potential leads to power law scale factor solution whereas, in Brans-Dicke theory, power law potential leads to power law solution of the scale factor.}. Also, instead of choosing $\omega_{\rm BD}$ and the exponent $q$ of the power law potential as the model parameters, we consider the power law exponent $n$ of the scale factor in the Brans-Dicke frame as our first model parameter as it determines the effective nature of the system. In order to study the viability of the scale factor, i.e. whether it may lead to scale-invariant spectra, we consider the conformally connected Einstein frame scale factor exponent $\alpha$ as our second model parameter as the perturbations are invariant under conformal transformation. And, the last model parameter is the same as before in Refs. \cite{Hrycyna:2013hla, Hrycyna:2013yia} as defined by the equation of state $w_m$ of the additional barotropic fluid. We study the attractor behavior and obtain the required condition for the stability of those solutions, mainly the contracting Universe solution. In order to understand the behavior, we provide an example of matter contracting Universe that leads to near scale-invariant spectra and investigate the system in the vicinity of the fixed point. We show that in this region, the system behaves like the Universe contains three different types of matter. Amongst them, with time, two decays down and the leading order solution survives. Since, we concentrate on the contracting phase of a bouncing scenario, we do not analyze the viability of the model through the constraints on the effective coupling constant, i.e.,  the {\it Newtonian constant}, $G$. We reserve ourselves to study the bouncing model across the bouncing phase in the future works.

The work is arranged in the following way. In the next section, we briefly introduce the Brans-Dicke theory with the corresponding field equations. 
In section \ref{Sec:PowerLawCosmo}, we formulate the power law potential and the corresponding power law scale factor solution in the absence of an additional matter. We also obtain the relations among the Brans-Dicke parameter and the power law potential with the power law scale factor solution in the Brans-Dicke frame and the same in the Einstein frame. In section \ref{Sec:PhaseSpace}, we express all cosmological equations in terms of the dimensionless quantities in the presence of an additional barotropic fluid.  By briefly introducing the two-dimensional phase space, we also obtain the fixed points in the presence of an additional barotropic matter. Then, we study the stability of those fixed points and obtain the general condition of stability. We identify the desired power law solution from the fixed points and show that, contraction can, indeed, lead to an attractor solution. In section \ref{Sec:MatterContraction}, we also provide an ideal example of matter contracting Universe which may lead to nearly scale-invariant spectra and obtain the solution near the vicinity of the desired power law fixed point. In the end, we conclude our work with the future outlook.

In this work, we use the $(-, +, +, +)$ metric signature convention. $\nabla_\mu$ and $\Box$ are defined as the covariant derivative and the d'Alembertian operator in curved spacetime, respectively.  Also, an overdot is defined as the partial derivative with respect to the cosmic time, $\sim\frac{\partial }{\partial t}$.

\section{The model}\label{Sec:The Model}

The action for the Brans-Dicke theory \cite{Brans-Dicke1961} is given by

\begin{eqnarray}\label{Eq:ActionBransDickeGeneral}
	\mathcal{S}_{\text{BD}} = \frac{1}{2}\int {\rm d}^4 {\bf x} \,\sqrt{-g}\left(\varphi R - \frac{\omega_{\rm BD}}{\varphi}\,g^{\mu \nu} \partial_\mu \varphi \partial_\nu \varphi - 2\,V(\varphi)\right) + 8\pi \int {\rm d}^4 {\bf x} \,\sqrt{-g}\,\mathcal{L}_{\rm m},
\end{eqnarray}
where, $R$ is the Ricci scalar, $\omega_{\rm BD}$ is the Brans-Dicke parameter, $\varphi$ is the scalar field, $V(\varphi)$ is the corresponding scalar field potential and the second part of the action describes the additional matter action. Varying the action (\ref{Eq:ActionBransDickeGeneral}) with respect to the metric $g_{\mu \nu}$ provides the metric field equation

\begin{eqnarray}\label{Eq:EoMBDGengmunu}
	\varphi\left(R_{\mu \nu} - \frac{1}{2}g_{\mu \nu}R\right) &=& \frac{\omega_{\rm BD}}{\varphi} \left(\nabla_\mu \varphi \nabla_\nu \varphi - \frac{1}{2} g_{\mu \nu} g^{\alpha \beta}\nabla_\alpha \varphi \nabla_\beta \varphi \right) - g_{\mu \nu} V(\varphi)  \nonumber \\
	&&\quad \quad\quad-\left(g_{\mu \nu }\,\Box\varphi  - \nabla_{\mu \nu}\varphi \right) + 8\pi\,T_{\mu \nu}^{(m)},
\end{eqnarray}
where,
\begin{eqnarray}
T_{\mu \nu}^{ (m)} \equiv-\frac{2}{\sqrt{-g}}\frac{\delta}{\delta g^{\mu \nu}}\left(\sqrt{-g}\mathcal{L}_{\rm m}\right)
\end{eqnarray}
is the energy momentum tensor of the barotropic fluid. The trace of this equation (\ref{Eq:EoMBDGengmunu}) takes the form
\begin{eqnarray}\label{Eq:EoMBDGenTrace}
	R  = \frac{\omega_{\rm BD}}{\varphi^2} g^{\alpha\beta}\nabla_\alpha\varphi\nabla_\beta\varphi + \frac{4}{\varphi}V(\varphi) + \frac{3}{\varphi}\Box\varphi - \frac{8\pi}{\varphi}T^{ (m)},
\end{eqnarray}
where, $T^{(m)}$ is the trace of the energy momentum tensor $T^{(m)}_{\mu \nu}$.  Conservation of the energy momentum tensor of the additional matter leads to \begin{eqnarray}\label{Eq:COEMT}
\nabla_\mu T^{(m)\,\mu}_{~~~~\,\nu} = 0.
\end{eqnarray}
Also, variation of the action (\ref{Eq:ActionBransDickeGeneral}) with respect to the scalar field gives
\begin{eqnarray}\label{Eq:EoMBDGenScalarField}
	\Box\varphi  = \frac{1}{2\varphi} g^{\alpha\beta} \nabla_\alpha\varphi\nabla_\beta\varphi - \frac{\varphi}{2 \omega_{\rm BD}}\left(R - 2V_{\varphi}(\varphi)\right).
\end{eqnarray}

\noindent $V_\varphi$ is defined as the partial derivative with respect to the scalar field, i.e.,  $\frac{\partial V}{\partial\varphi}$. These three equations (\ref{Eq:EoMBDGengmunu}), (\ref{Eq:EoMBDGenTrace}) and (\ref{Eq:EoMBDGenScalarField}) in the absence of the additional matter are, however, not independent. Due to the constrained nature of gravity, only two of them are independent.

In our case, we consider the Friedmann-Robertson-Walker (FRW) homogeneous and isotropic Universe with the line element
\begin{eqnarray}\label{Eq:FRWLineElement}
	{\rm d}s^2 = -{\rm d}t^2 + a^2(t)\,{\rm d}{\bf x}^2
\end{eqnarray}
where $t$ is the cosmic time and $a(t)$ is the scale factor of the Universe. We consider the additional matter as the barotropic fluid which is described by the equation of state $w_m = \frac{p_m}{\rho_m}$. $p_m$ and $\rho_m$ are the pressure and energy density of the fluid, respectively and are defined as
\begin{eqnarray}\label{Eq:PmRhom}
	\rho_m \equiv - T^0_0,\quad T^i_j \equiv p_m\delta^i_j.
\end{eqnarray}
Using the line element (\ref{Eq:FRWLineElement}), the equations (\ref{Eq:EoMBDGengmunu}), (\ref{Eq:EoMBDGenTrace}) and (\ref{Eq:EoMBDGenScalarField}) can be reduced further. The 0-0 component of the equation (\ref{Eq:EoMBDGengmunu}) becomes the energy constrained equation with the form
\begin{eqnarray}\label{Eq:0-0Hom}
	3H^2 = \frac{\omega_{\rm BD}}{2}\frac{\dot{\varphi}^2}{\varphi^2} + \frac{V(\varphi)}{\varphi} - 3H\frac{\dot{\varphi}}{\varphi} + \frac{8\pi}{\varphi} \rho_m, \quad H\,\equiv \frac{1}{a}\frac{{\rm d} a(t)}{{\rm d}t}.
\end{eqnarray}
$H(t)$ is the Hubble parameter. The trace equation (\ref{Eq:EoMBDGenTrace}) in FRW Universe becomes the acceleration equation
\begin{eqnarray}\label{Eq:AcclEq}
	\dot{H} = - \frac{\omega_{\rm BD}}{2}\frac{\dot{\varphi}^2}{\varphi^2} - \frac{1}{3 + 2 \omega_{\rm BD}}\frac{2 V(\varphi) - \varphi V_\varphi(\varphi)}{\varphi} + 2 H \frac{\dot{\varphi}}{\varphi} -\frac{8\pi\rho_m}{\varphi}\frac{2 + \omega_{\rm BD}(\left(1 + w_m\right))}{3 + 2 \omega_{\rm BD}}
\end{eqnarray}
and the scalar field equation (\ref{Eq:EoMBDGenScalarField}) becomes
\begin{eqnarray}\label{Eq:ScEqHom}
	\ddot{\varphi} + 3 H \dot{\varphi}  = 2 \frac{2 V(\varphi) - \varphi V_\varphi(\varphi)}{3 + 2 \omega_{\rm BD}} + 8 \pi \rho_m\frac{1 - 3 w_m}{3 + 2\omega_{\rm BD}}.
\end{eqnarray}
Finally, the conservation of the energy-momentum tensor of the additional fluid (\ref{Eq:COEMT}) leads to
\begin{eqnarray}\label{Eq:EoMBar}
\dot{\rho}_m + 3\,H\,(\rho_m + P_m) = 0.
\end{eqnarray}
In the next section, using these background equations, first, we will construct the model and then, in the following sections, we will examine the phase space behavior.

\section{Power law cosmology}\label{Sec:PowerLawCosmo}

As mentioned earlier, the equations (\ref{Eq:0-0Hom}), (\ref{Eq:AcclEq}), (\ref{Eq:EoMBDGenScalarField}) and (\ref{Eq:EoMBar}) are not independent and hence, we cannot obtain the solution for $a(t), \,\varphi$ and $V(\varphi)$, uniquely. In this work, we are interested in power law solution of the scale factor and  we consider the following form:
\begin{eqnarray}\label{Eq:PowerLawSc}
	a(\eta) = a_0 \left(\frac{\eta}{\eta_0}\right)^n, \quad \varphi = \varphi_0\left(\frac{\eta}{\eta_0}\right)^m, \quad V  = V_0 \left(\frac{\eta}{\eta_0}\right)^p,
\end{eqnarray}
where, $\eta$ is the conformal time and is related to cosmic time by the relation $\eta \equiv \int {\rm d} t/a(t)$. If we ignore the barotropic matter, using the above power law form (\ref{Eq:PowerLawSc}), it can be shown that there exist only two independent parameters as the number of independent equations  is two and everything else can be written in terms of those two variables. Therefore, in this case, we can express $\omega_{\rm BD}, V_0$ and $q$ solely in terms of $n$ and $m$ defined in the above equation. Using any two of (\ref{Eq:0-0Hom}), (\ref{Eq:AcclEq}) and (\ref{Eq:ScEqHom}), we obtain the solution as
\begin{eqnarray}\label{Eq:wbdV0PowerLaw}
	\omega_{\rm BD} = \frac{-m^2 + m + 2 m n + 2 n + 2 n^2}{m^2}, \quad V_0 = \frac{(-m + m^2 - 2 n + 4 m n + 4 n^2)\, \varphi_0}{2\, a_0^2} 
\end{eqnarray}
with 
\begin{eqnarray}
	p = m- 2 - 2n,
\end{eqnarray}
and
\begin{eqnarray}\label{Eq:Pot}
	V = V_0 \,\left(\frac{\varphi}{\varphi_0}\right)^q, \quad q = 1 - 2\left(\frac{1 + n}{m}\right).
\end{eqnarray}

Using the above relations, we are now able to construct any model which resembles power law scale factor. For example, $n = -1$ corresponds to pure de-Sitter universe in the Brans-Dicke theory, which can be achieved by the potential with $q = 1$, i.e., the potential is linear with $m \neq 0$. In our case, instead of using variable $m$, we choose a new variable $\alpha$ which is defined as

\begin{eqnarray}\label{Eq:IntroAlpha}
	m = 2\,(\alpha - n),
\end{eqnarray}
where $\alpha$ represents the exponent of the power law scale factor $a  \propto \eta^\alpha$ in the conformally connected Einstein frame. This can easily be verified as the corresponding conformal transformation from Brans-Dicke frame to Einstein frame is
\begin{eqnarray}
	g^{(\rm E)}_{ij} = \varphi\,g^{(\rm J)}_{ij} \quad
	 \Rightarrow \alpha = n + \frac{m}{2},
\end{eqnarray}
which is same as (\ref{Eq:IntroAlpha}). Replacing $m$ by $\alpha$ in (\ref{Eq:wbdV0PowerLaw}) and (\ref{Eq:Pot}), we obtain
\begin{eqnarray}\label{Eq:wbdq}
	\omega_{\rm BD} = \frac{-3\, (n - \alpha)^2 + \,\alpha(\alpha+ 1)}{2\, (n - \alpha)^2}, \quad q = 1 + \frac{n + 1}{n - \alpha},
\end{eqnarray}
 and 
 
 \begin{eqnarray}\label{Eq:PotAlpha}
 	V_0 = \frac{\alpha(2\alpha -1)\varphi_0}{a_0^2}.
 \end{eqnarray}
 
\noindent We will show in the next section that, the constant $V_0$ plays a crucial role in determining the phase space dynamics.

The above equations, however, do not hold true for some specific situations. For instance, consider the situation where $ m = 2 \,(\alpha - n) = 0$, i.e., $n = \alpha$. This implies that, the exponent of the power law scale factor in both the frames are identical. Then, for $n = \alpha \neq -1$, exponent of the power law potential, $q$ diverges and the solution deviates from the power law behavior. Therefore, in this work, we do not consider such cases where the scale factors in the two different conformally coupled frames are identical.

However, consider the special situation where the Universe is de-Sitter in the Einstein frame with $\alpha = -1$. In that case, $q = 2$ (assuming $n \neq \alpha$). As for the Brans-Dicke parameter, for $n \neq \alpha$, such condition leads to $\omega_{\rm BD} = -3/2$, as can be seen from (\ref{Eq:wbdq}). This choice always leads to pathologies \cite{Dabrowski:2005yn} as $n$ is undetermined. This can be averted if we choose $n = \alpha = -1$, i.e., in all conformally connected frames, the solutions are de-Sitter. In this case, however, $\omega_{\rm BD}$  is undetermined and  the solution around the fixed point acts as a global attractor for $w_m > -1$. This case has already been studied in some detail earlier (see Refs. \cite{Hrycyna:2013hla, Hrycyna:2013yia}). Note that, since $\omega_{\rm BD}$ is undetermined, the relation remains true for $f(R)$ Starobinsky model with $\omega_{\rm BD} = 0$ as well \cite{STAROBINSKY1982175, Starobinsky:1987zz}. In this work, instead of assuming such a specific choice, we generalize the study for arbitrary power law behavior.

\section{Phase space analysis}\label{Sec:PhaseSpace}

In the previous section, we had constructed the model and solved the equation in the absence of the additional matter. In this section, to study the behavior of the phase space, we consider the additional barotropic matter with the equations of motion (\ref{Eq:0-0Hom}), (\ref{Eq:AcclEq}), (\ref{Eq:ScEqHom}) and (\ref{Eq:EoMBar}). Instead of using quantities with dimensions, we can express and re-write the field equations in terms of dimensionless quantities. These quantities are

\begin{eqnarray}\label{Eq:DefXY}
	x  \equiv \frac{\dot{\varphi}}{H \,\varphi}, \quad y \equiv\frac{1}{H}\sqrt{\frac{V(\varphi)}{3\varphi}} = \frac{1}{H}\sqrt{\frac{V_0}{3\,\varphi_0{}^q}} \,\varphi^{(q -1)/2}.
\end{eqnarray}

Then, using these quantities, the energy constrained equation (\ref{Eq:0-0Hom}) becomes a constraint equation in the phase space and  can be written as
\begin{eqnarray}\label{Eq:Constrainedxy}
	\Omega_m \equiv \frac{8\pi \rho_m}{3 \varphi H^2} = 1 + x - \frac{\omega_{\rm BD} }{6} x^2 - y^2,
\end{eqnarray}
where, $\Omega_m$ is the energy density fraction of the barotropic fluid. Similarly, the acceleration equation (\ref{Eq:AcclEq}) can be expressed as
\begin{eqnarray}\label{Eq:ConstrainedAcc}
	\frac{\dot{H}}{H^2} = 2x - \frac{\omega_{\rm BD}}{2}x^2 - \frac{3(2 - q)}{3 + 2 \omega_{\rm BD}}y^2 - 3 (1 + x - \frac{\omega_{\rm BD}}{6}x^2 - y^2)\frac{2 + \omega_{\rm BD}(1 + w_m)}{3 + 2 \omega_{\rm BD}}.
\end{eqnarray}

Defining equation of state of the system in a non-minimally coupled frame is not unique as the energy momentum tensor is coupled to the other field. Instead, we  re-arrange and re-write field equations in a similar fashion in Einstein frame, i.e., $G_{\mu \nu} = T^{(\rm eff)}_{\mu \nu}(g_{\alpha\beta}, \varphi)$, where $G_{\mu \nu}$ is the Einstein tensor and we can define $T^{(\rm eff)}_{\mu \nu}(g_{\alpha\beta}, \varphi)$ as the effective energy momentum tensor.  This ensures that the definition of effective equation state becomes identical to the definition in Einstein frame and it depends only on the scale factor in the respective frame as
\begin{eqnarray}\label{Eq:EffectiveEOS}
w_{\rm eff} = -1 - \frac{2}{3}\frac{\dot{H}}{H^2}.
\end{eqnarray}

As discussed earlier, without the additional matter, due to diffeomorphism, the degrees of freedom of the system is one. Hence, the phase space is one dimensional. With the additional matter, the degrees of freedom in the system increases and the phase space dimension becomes two. In order to study the dynamics, instead of using cosmic or conformal time, we use the `e-fold' time convention with $N \equiv {\rm ln}\left(a(t)\right) = \int H(t){\rm d}t$. Using equations (\ref{Eq:0-0Hom}), (\ref{Eq:AcclEq}) and (\ref{Eq:EoMBDGenScalarField}), we obtain the evolution equations of $x$ and $y$ as	
\begin{eqnarray}\label{Eq:EoMX}
	\frac{{\rm d} x}{{\rm d}N} &=& - 3x - x^2 - x\,\frac{\dot{H}}{H^2} + \frac{6(2 - q)}{ 3 + 2 \omega_{\rm BD}}y^2  + 3 (1 + x - \frac{\omega_{\rm BD}}{6}x^2 - y^2)\frac{1 - 3 w_m}{3 + 2 \omega_{\rm BD}}, \\
	\label{Eq:EoMY}
	\frac{{\rm d} y}{{\rm d}N} &=& \frac{(q - 1)}{2} x y - y\,\frac{\dot{H}}{H^2} .
\end{eqnarray}

While the direction of $N$ is positive in an expanding Universe, contracting Universe ensures $N$ to go in the negative direction.  The above two equations (\ref{Eq:EoMX}) and (\ref{Eq:EoMY}) along with (\ref{Eq:ConstrainedAcc}) and (\ref{Eq:Constrainedxy}) govern the whole phase space dynamics. In the following sections, we will use these equations to study the dynamical behavior of the system.

\subsection{Fixed points}

Now, using equations (\ref{Eq:EoMX}) and (\ref{Eq:EoMY}) along with (\ref{Eq:ConstrainedAcc}), we can find the critical or the fixed points of the system. These points define the exact solutions of the system, i.e., the {\it desired effective leading order} solutions. These points can be obtained by equating (\ref{Eq:EoMX}) and (\ref{Eq:EoMY}) to be zero, i.e., the velocities of $x$ and $y$ vanish at these points. This corresponds to seven such points:
	 \begin{eqnarray}\label{Eq:FixedPoint1}
		1.\quad x_1^* &=& \frac{4 - 2 q }{ 1 +  q + 2 \omega_{\rm BD}}, \quad y_1^* = -\frac{1}{\sqrt{3}}\frac{\sqrt{(5 + 4 q + 6 \omega_{\rm BD} - q^2)( 3 + 2 \omega_{\rm BD})}}{1 +  q + 2 \omega_{\rm BD}} \\
		\label{Eq:FixedPoint2}
	2. \quad x_2^* &=& \frac{4 - 2 q }{ 1 +  q + 2 \omega_{\rm BD}}, \quad y_2^* = \frac{1}{\sqrt{3}}\frac{\sqrt{(5 + 4 q + 6 \omega_{\rm BD} - q^2)( 3 + 2 \omega_{\rm BD})}}{1 +  q + 2 \omega_{\rm BD}}
	\end{eqnarray}

Only these two points always describe the desired behavior of the system that we choose, i.e., the scale factor in Brans-Dicke frame is $\sim \eta^n$ whereas, in Einstein frame, it is $\sim \eta^\alpha$. This can indeed be verified with the help of the effective equation of state (\ref{Eq:EffectiveEOS}). We will discuss it in detail in the next section.

	\begin{eqnarray}
	\label{Eq:FixedPoint3}
	3. \quad x_3^* &=& \frac{3 - \sqrt{9 + 6 \omega_{\rm BD}}}{\omega_{\rm BD}}, \quad y_3^* = 0 \\
	\label{Eq:FixedPoint4}
	4. \quad x_4^* &=& \frac{3 + \sqrt{9 + 6 \omega_{\rm BD}}}{\omega_{\rm BD}}, \quad y_4^* = 0
	\end{eqnarray}

These two fixed points are a special case as they do not depend on $q$. This is the reason we get the same form of these fixed points even in $q = 2$ case like in Ref. \cite{Hrycyna:2013yia}. These points are the solution for zero potential, i.e., $V_0 = 0$ (whereas, the first two points are valid for non-zero finite potential) and completely governed by the kinetic part of the action (\ref{Eq:ActionBransDickeGeneral}). As a consequence, around or at these fixed points, the system behaves differently, i.e., the Universe is described by different scale factor solutions. Again, this can be verified from the effective equation of state (\ref{Eq:EffectiveEOS}).

These four points are independent of the additional matter defined by Lagrangian density $\mathcal{L}_m$ in (\ref{Eq:ActionBransDickeGeneral}) as they are independent of $w_m$. In fact, we will show in the next section that these points correspond to $\Omega_m = \rho_m = 0$, i.e., the solutions (by solving the fixed points) exist only in the absence of the additional matter. The last three, however, are not. Therefore, the last three solutions represent the dynamics of the scalar field coupled with the additional barotropic matter. They are

	\begin{eqnarray}
		\label{Eq:FixedPoint5}
	5. \quad	x_5^* &=& \frac{-3 (1 + w_m)}{q}, \quad y_5^* =  -\frac{\sqrt{\begin{aligned}
				&q^2 (1 - 3 w_m) - 3 (1 + 2 \omega_{\rm BD})(-1 + \omega_{\rm BD} (w_m - 1)) \nonumber\\
				&\times(1 + w_m) + q (4 + \omega_{\rm BD}(5 - 6w_m - 3 w_m^2))
				\end{aligned}}}{\sqrt{2}  \sqrt{1 + q + 2 \omega_{\rm BD}}} \nonumber\\
		\\
		\label{Eq:FixedPoint6}
	6. \quad	x_6^* &=& \frac{-3 (1 + w_m)}{q}, \quad y_6^* =  \frac{\sqrt{\begin{aligned}
				&q^2 (1 - 3 w_m) - 3 (1 + 2 \omega_{\rm BD})(-1 + \omega_{\rm BD} (w_m - 1)) \nonumber\\
				&\times(1 + w_m) + q (4 + \omega_{\rm BD}(5 - 6w_m - 3 w_m^2))
				\end{aligned}}}{\sqrt{2}  \sqrt{1 + q + 2 \omega_{\rm BD}}}\nonumber\\
		\\
		\label{Eq:FixedPoint7}
	7. \quad	x_7^* &=& \frac{1 - 3 w_m}{1 + \omega_{\rm BD} - \omega_{\rm BD}w_m}, \quad y_7^* = 0
	\end{eqnarray}

The fifth (\ref{Eq:FixedPoint5}) and sixth (\ref{Eq:FixedPoint6}) fixed points are barotropic fluid dominated solution in the presence of the scalar field with finite potential. As we will show later is that, most of the time, these solutions are not physical as the matter energy density fraction $\Omega_m < 0$. The last solution (\ref{Eq:FixedPoint7}) is also the additional matter dominated solution coupled with the scalar field, however, with zero potential. Again, just like fifth and sixth solutions, this solution is also not physical for some values of $\omega_{\rm BD}, q$ and $w_m$.

One immediate result that can easily be verified from the above expressions is that, by directly substituting $q = 2$, we can recover the results of Ref. \cite{Hrycyna:2013yia}. In this case, as we already have discussed, the model corresponds to $n = \alpha = -1$, i.e., in both the frames, the desired scale  factor is de-Sitter in the presence of finite potential. Those de-Sitter states are represented by the first two points (\ref{Eq:FixedPoint1}) and (\ref{Eq:FixedPoint2}).

Now, in order to study the stability of these fixed points, we need to linearize the equations (\ref{Eq:EoMX}) and (\ref{Eq:EoMY}) as
\begin{eqnarray}\label{Eq:LinearizedEq}
\left(\begin{aligned}
& \delta x^\prime\\
& \delta y^\prime
\end{aligned}\right) = \left(\begin{aligned}
\frac{\partial A(x, y)}{\partial x}\Big|_* && \frac{\partial A(x, y)}{\partial y}\Big|_* \\
\frac{\partial B(x, y)}{\partial x}\Big|_* && \frac{\partial B(x, y)}{\partial y}\Big|_*
\end{aligned}\right)\left(\begin{aligned}
\delta x &\\\delta y &
\end{aligned}\right),
\end{eqnarray}
where, $A(x, y)$ and $B(x, y)$ are the right-hand side of (\ref{Eq:EoMX}) and (\ref{Eq:EoMY}), respectively. $|_*$ denotes the value at the fixed point. By linearizing equations, we assume that we are studying the stability condition in the vicinity of the fixed points, i.e., the deviation from the fixed points are very small.

In order to check the stability of the fixed points, we need to evaluate the eigenvalues of this matrix. If both the eigenvalues are negative (positive) in an expanding (contracting) Universe, then $\delta x$ and $\delta y$ approach zero as $N$ approaches $\infty\, (-\infty)$. This implies that the deviation from the actual trajectory reduces with time and the new trajectory returns to the original trajectory asymptotically in time. In other words, both eigenvalues being negative (positive) in an expanding (contracting) Universe  implies that the fixed point is stable and the corresponding solution is an attractor solution.

In the following sections, using this definition, we will determine the general condition for the stability of the fixed points. Also, in order to understand the behavior of the system, we will evaluate the corresponding effective equation of state (\ref{Eq:EffectiveEOS}) as well as the fractional matter energy density (\ref{Eq:Constrainedxy}) at those points.

\subsection{The first and the second fixed points}

Consider the first two fixed points given in (\ref{Eq:FixedPoint1}) and (\ref{Eq:FixedPoint2}). For these points, we can evaluate the eigenvalues for the linearized matrix defined in (\ref{Eq:LinearizedEq}). Along with the eigenvalues, we can calculate the effective equation of state (\ref{Eq:EffectiveEOS}) as well as the energy density fraction for the barotropic matter (\ref{Eq:Constrainedxy}). In terms of $q, \, \omega_{\rm BD}$ and $w_m$, these are given in table \ref{Tab:Value12first}.

\begin{table}[H]
\begin{center}
	\begin{tabular}{|c|c|}
		\hline
		&\\
		$x_{(1, 2)}^*$ &  $\frac{4 - 2 q }{ 1 +  q + 2 \omega_{\rm BD}}$ \\[10pt]
		\hline
		&\\
		$ y^*_{(1, 2)}$ & $\mp \frac{1}{\sqrt{3}}\frac{\sqrt{(5 + 4 q + 6 \omega_{\rm BD} - q^2)( 3 + 2 \omega_{\rm BD})}}{1 +  q + 2 \omega_{\rm BD}} $ \\[10pt]
		\hline &\\
		$\lambda_1$ & $\frac{ q^2- 4 q - 5 - 6 \omega_{\rm BD}}{1 + q + 2 \omega_{\rm BD}}$ \\[10pt]
		\hline &\\
		$\lambda_2$ & $\frac{2 q^2 - q(7 + 3w_m) - 3 (1 + 2 \omega_{\rm BD})(1 + w_m)}{1 + q + 2 \omega_{\rm BD}}$ \\[10pt]
		\hline &\\
		$w_{\rm eff}$ & $\frac{2 q^2 - 9 q + 1 - 6 \omega_{\rm BD}}{3(1 + q + 2 \omega_{\rm BD})}$\\[10pt]
		\hline &\\
		$\Omega_m$ & $0$ \\[5pt]
		\hline
	\end{tabular}
	\caption{Behavior of the first and second fixed points in terms of the model parameters $q,\, \omega_{\rm BD}$ and $w_m$.}
	\label{Tab:Value12first}
\end{center}
\end{table}

As one can see, at these points, $\Omega_m$ is precisely equal to zero. These correspond to $\rho_m = 0$ and can be verified from the equation (\ref{Eq:Constrainedxy}). This implies that the solution is only valid in the absence of additional matter and any deviation from these points represents a finite value of $\rho_m$, i.e., the presence of additional matter.

As we have mentioned earlier, instead of using $\omega_{\rm BD}$ and $q$ as the model parameters, in this work, we work with $\alpha,$ the Einstein frame scale factor parameter and $n$, the Brans-Dicke frame scale factor parameter as the new model parameters. Table \ref{Tab:Value12second} contains the fixed points as well as the eigenvalues, effective equation of state and the energy density fraction in terms of these new model parameters.

\begin{table}[H]
\begin{center}
	\begin{tabular}{|c|c|}
		\hline
		&\\
		$x_{(1, 2)}^*$ &  $-2 + \frac{2\alpha}{n}$ \\[10pt]
		\hline
		&\\
		$ y^*_{(1, 2)}$ & $\mp \frac{\sqrt{\alpha(2\alpha-1)}}{\sqrt{3}n} $ \\[10pt]
		\hline &\\
		$\lambda_1$ & $\frac{1 - 2\alpha}{n}$ \\[10pt]
		\hline &\\
		$\lambda_2$ & $\frac{ 2 - 2\alpha + n(1 - 3w_m)}{n}$ \\[10pt]
		\hline &\\
		$w_{\rm eff}$ & $\frac{2 - n}{3\,n}$\\[10pt]
		\hline &\\
		$\Omega_m$ & $0$ \\[5pt]
		\hline
	\end{tabular}
	\caption{Behavior of the first and second fixed points in terms of the model parameters $n,\, \alpha$ and $w_m$.}
	\label{Tab:Value12second}
\end{center}

\end{table}

Let us first discuss the nature of the fixed points. The $\mp$ sign in the $y^*$ value denotes the two different phases of the same scale factor: expanding $(+)$ and contracting phase $(-)$. Now consider the effective equation of state in table \ref{Tab:Value12second} along with the equation (\ref{Eq:EffectiveEOS}). This implies that the effective scale factor in the Brans-Dicke frame is indeed $\sim \eta^n$ which is how we construct the model and is completely independent of $\alpha$ and $w_m$. Therefore, only these two points always correspond to the solution of how we construct the model.

Now consider the corresponding eigenvalues. The first fixed point (contraction) (\ref{Eq:FixedPoint1}) is a stable node if and only if
\begin{eqnarray}\label{Eq:12lamda1}
\lambda_1 = \frac{1 - 2\alpha}{n} \,> 0, \quad \lambda_2 = \frac{2 - 2\alpha + n(1 - 3w_m)}{n} \,> 0
\end{eqnarray}
whereas, the second fixed point (expansion) (\ref{Eq:FixedPoint2}) is stable if and only if 
\begin{eqnarray}\label{Eq:12lamda2}
\lambda_1 = \frac{1 - 2\alpha}{n} \,< 0, \quad \lambda_2 = \frac{2 - 2\alpha + n(1 - 3w_m)}{n} \,< 0.
\end{eqnarray}

Let us first understand what the stability of these fixed points signifies. As I mentioned earlier, these points correspond to $\rho_m = 0$, i.e., the Universe contains only the scalar field $\varphi$. The deviation from the fixed points implies that the Universe contains another matter with finite energy density $\rho_m$. Then, the stability of those points implies that, whether the solution with the additional matter will asymptotically track the solution without the additional matter. In other words, if the points are stable, then, even the presence of additional matter leads to the same fixed point solution, asymptotically,  which is the scalar field dominated solution.

The results (\ref{Eq:12lamda1}) and (\ref{Eq:12lamda2}) are {\it surprising and interesting}. The eigenvalues not only depend on  $\alpha$ but also depend on $n$. This means that, after conformal transformation, stability condition in the Einstein frame does not guarantee the stability in the conformally connected Brans-Dicke frame.

To go about understanding this, we will consider three cases. In each of these cases, we will fix any two model parameters of $(n, \alpha, w_m)$ in the parameter space and observe how the other variable determines the stability.

\paragraph{Case I:} In this case, we first fix $n$ and $w_m$ and keep $\alpha$ arbitrary in the parameter space, i.e.,  $(n, \alpha, w_m) = (2, \alpha, 1/3)$. The corresponding eigenvalues are
\begin{eqnarray}\label{Eq:2-alpha-1/3}
	\lambda_1 = \frac{1}{2} - \alpha, \quad \lambda_2 = 1 - \alpha,
\end{eqnarray}
which implies that for any value of $\alpha < 1/2$, both of the eigenvalues are positive.  $ n = 2$ implies that in the Brans-Dicke frame, i.e., in our physical frame, {\it the Universe is matter dominated} and $w_m = 1/3$ signifies that {\it the additional matter is relativistic.} For $\alpha < 1/2$, the positive signs of the eigenvalues confirm that the point is not stable if the Universe is expanding, i.e., the second fixed point  (\ref{Eq:FixedPoint2}) is not stable. However, the first fixed point (\ref{Eq:FixedPoint1}) corresponds to the contracting Universe and thus the point is stable. In other words, if the Universe contains relativistic particles with the scalar field defined by the potential and the Brans-Dicke parameter corresponding to $n = 2, \,\alpha < 1/2$, then eventually the solution leads to the solution without the additional matter, asymptotically. This result is {\it compelling} as we always commonly think of contracting Universe (except ekpyrosis \cite{Levy:2015awa}) as a non-attractor. This result clearly is in contradiction to that and proves that {\it contracting Universe can also be an attractor}. This is the main result of our work. In section \ref{Sec:MatterContraction}, we will discuss this case in detail by providing the approximate solution in presence of the additional matter.

Similarly, for $\alpha > 1$, both eigenvalues are negative and the expandin7g Universe solution (\ref{Eq:FixedPoint2}) is stable.

\paragraph{Case II:} In this case, we keep $n$ to be arbitrary in the parameter space, i.e., $(n, \alpha, w_m) = (n, 2, 1/3)$. This implies, in the Einstein frame, the Universe is matter dominated and in the Brans-Dicke frame, the additional matter is relativistic. In this case, the eigenvalues take the form

\begin{eqnarray}
\lambda_1 = -\frac{3}{n}, \quad \lambda_2 = - \frac{2}{n}.
\end{eqnarray}

This implies that, for any value of $n$ being negative, the first fixed point (\ref{Eq:FixedPoint1}), i.e., the contracting Universe solution is stable whereas, if $n$ is positive, then the second point (\ref{Eq:FixedPoint2}), i.e., the expanding Universe solution is stable.

\paragraph{Case III:} In this case, we keep $w_m$ to be arbitrary to determine the role of the barotropic fluid in determining the dynamical behavior of the system. For instance, for $(n, \alpha, w_m) = (2, -3/2, w_m)$,  the eigenvalues are
\begin{eqnarray}\label{Eq:2-3/2-wm}
	\lambda_1 = 2, \quad \lambda_2 = \frac{7 - 6 w_m}{2},
\end{eqnarray} 
which implies that the first fixed point (\ref{Eq:FixedPoint1}) is only stable if $w_m < 7/6$. This means that, if the Universe contains matter with the equation of state $w_m < 7/6$, the solution tracks the `solution without the additional matter' in time. However, if the equation of state of the additional matter is $\geq 7/6$, the corresponding solution will not track the actual solution, i.e., the solution becomes a non-attractor.

\paragraph{Effect of anisotropy}

Note that, in the most cases of bouncing solutions, the energy density due small anisotropy  grows heavily with time. This, in turn, can break the homogeneous and isotropic background, leading to Kasner/mixmaster chaos. The energy density due to tiny anisotropy behaves as $\rho_m \sim a^{-6}$. This corresponds to the equation of state of $w_m = 1. $ Therefore, by using the model parameter $w_m =1$, we can investigate the stability of the system in the presence of small anisotropy. For instance, if we consider the model parameters $(n, \alpha, w_m)$  as $(2, \alpha, 1)$, we obtain the eigenvalues corresponding to the fixed points \eqref{Eq:FixedPoint1} and \eqref{Eq:FixedPoint2}
	\begin{eqnarray}
	\lambda_{1} = \frac{1}{2} - \alpha, \quad \lambda_{2} = - (1 + \alpha).
	\end{eqnarray}
	
\noindent Hence, the matter contracting solution \eqref{Eq:FixedPoint1} being an attractor in the presence of anisotropy requires $\alpha < -1.$ This implies that, the energy density due to the presence of tiny anisotropy decays for $(2, \alpha < -1, 1)$ and therefore do not lead to Kasner/mixmaster chaos.

\subsection{The third and the fourth fixed points}

Let us discuss the next two points (\ref{Eq:FixedPoint3}) and (\ref{Eq:FixedPoint4}). The eigenvalues, effective equation of state and the energy fraction of the matter corresponding to these points are given in the table \ref{Tab:Value34} below:

\begin{table}[H]
	\begin{center}
		\begin{tabular}{|c|c|}
			\hline
			&\\
			$x_{(3, 4)}^*$ &  $\frac{3 \mp \sqrt{9 + 6 \omega_{\rm BD}}}{\omega_{\rm BD}}$ \\[10pt]
			\hline
			&\\
			$ y^*_{(3, 4)}$ & $0$ \\[10pt]
			\hline &\\
			$\lambda_1$ & $3 (1- w_m) + x^*_{(3, 4)}$ \\[10pt]
			\hline &\\
			$\lambda_2$ & $3 + \frac{3}{2}\,x^*_{(3, 4)}$ \\[10pt]
			\hline &\\
			$w_{\rm eff}$ & $1 + \frac{2}{3}\, x^*_{(3, 4)}$\\[10pt]
			\hline &\\
			$\Omega_m$ & $0$ \\[5pt]
			\hline
		\end{tabular}
		\caption{Behavior of the third and fourth fixed points in terms of the model parameters $q,\, \omega_{\rm BD}$ and $w_m$.}
		\label{Tab:Value34}
	\end{center}
	
\end{table}

As usual, these points also represents $\Omega_{m} = \rho_m = 0$, i.e., there is no additional matter present in the system. These points are  also the special points as $y^* = 0$ confirming that $V_0 = \tilde{V}_0 = 0$, i.e.,  the potential term in (\ref{Eq:ActionBransDickeGeneral}) vanishes. Clearly, in this case (and for the following points), the Brans-Dicke  and the conformally connected Einstein Universe do not evolve as $ \sim \eta^n$ and $\sim \eta^\alpha$, respectively. In order to determine the evolution of the Universe, we use the equation (\ref{Eq:PotAlpha}) and we obtain
\begin{eqnarray}\label{Eq:34alpha}
	\tilde{\alpha} = 0, \,\frac{1}{2}.
\end{eqnarray}

$\tilde{\alpha}$ (is not related to $\alpha$) defines how the system behaves in the Einstein frame ($a(\eta)\sim \eta^{\tilde{\alpha}}$). In this case, only $\omega_{\rm BD}$ determines the scale factor $a \sim \eta^{\tilde{n}}$  in the Brans-Dicke frame as the potential vanishes. The given values of $n$ and $\alpha$, in this case, only determine the value of $\omega_{\rm BD}$ which in turn determines the value $\tilde{n}$, the effective scale factor in the Brans-Dicke frame. The solution $\tilde{\alpha} = 0$ in (\ref{Eq:34alpha}) corresponds to $\omega_{\rm BD} = -3/2$ for $n \neq \alpha$ which can be seen from (\ref{Eq:wbdV0PowerLaw}). Therefore, $\tilde{\alpha} = 0$ solution is ruled out. For $\tilde{\alpha} = 1/2$, the Universe is governed by the scalar field defined by the effective equation of state given in the table \ref{Tab:Value34}. Let us consider the previous case of $(n, \alpha, w_m) = (2, -3/2, 1/3)$. Using these variables, we can evaluate $\omega_{\rm BD}$ using (\ref{Eq:wbdV0PowerLaw}) and it becomes $-72/49$. It implies that the system is defined by the zero potential with $\omega_{\rm BD} = -72/49$ in the absence of the additional matter. Therefore, 

\begin{eqnarray}
	\omega_{\rm BD}|_{\alpha \rightarrow 1/2} &=& -\frac{72}{49} \nonumber \\
	\Rightarrow \tilde{n} &=& -3, \, 4.
\end{eqnarray}

As it can be seen from (\ref{Eq:DefXY}), $x$ is defined as $2(\tilde{\alpha} - \tilde{n})/\tilde{n}$. Hence, the corresponding values of $\tilde{n} = -3, 4$ are $x = -7/3$ and $-7/4$, respectively. This result can be verified directly from the above table \ref{Tab:Value34}. Also, it is clear that the third fixed point (\ref{Eq:FixedPoint3}) corresponds to $\tilde{n} = 4$ solution where as the fourth fixed (\ref{Eq:FixedPoint4}) corresponds to $\tilde{n} = -3$. The equation of state for (\ref{Eq:FixedPoint3})  and (\ref{Eq:FixedPoint4}) are
\begin{eqnarray}
w_{\rm eff} = - \frac{1}{6}, \, -\frac{5}{9}
\end{eqnarray}
respectively. Again, this result can be verified from the above table \ref{Tab:Value34}.

The eigenvalues corresponding to the third point (\ref{Eq:FixedPoint3}) for $w_m = 1/3$ are
\begin{eqnarray}\label{Eq:Eig3}
\lambda_1 = \frac{1}{4}, \quad \lambda_2 =  \frac{1}{2},
\end{eqnarray}

This implies that the fixed point (\ref{Eq:FixedPoint3}) is only stable if the Universe is contracting.  Similarly, eigenvalues for the fourth point (\ref{Eq:FixedPoint4}) for $w_m = 1/3$ are 
\begin{eqnarray}\label{Eq:Eig4}
\lambda_1 = -\frac{1}{3}, \quad \lambda_2 =  -\frac{1}{3}
\end{eqnarray}
Therefore, the fixed point is only stable if the Universe is expanding.

Since these two fixed points are the solution for zero potential, this analysis is valid for the same fixed points in \cite{Hrycyna:2013yia}.

\subsection{Last three fixed points}

The fifth (\ref{Eq:FixedPoint5}) and sixth (\ref{Eq:FixedPoint6}) fixed points are the solutions due to the presence of an additional matter field which is minimally coupled to the scalar field. In terms of $(n, \alpha, w_m)$, these fixed points, as well as their eigenvalues, effective equation of state and energy fraction of the additional matter, are expressed in the table \ref{Tab:Value56}:

\begin{table}[H]
	\begin{center}
		\begin{tabular}{|c|c|}
			\hline
			&\\
			$x_{(5, 6)}^*$ &  $-\frac{3 (1 + w_m)(n - \alpha)}{1 + 2n - \alpha}$ \\[10pt]
			\hline
			&\\
			$ y^*_{(5, 6)}$ & $\mp \frac{\sqrt{n^2(1 - 3w_m)^2  + 2n(1 - 3 w_m)(1 + 3w_m \alpha)
					+\alpha(1 + 6 w_m + 2\alpha  + w_m^2(6\alpha-3)) 
			}}{2(\alpha - 1 - 2n)}$ \\[10pt]
			\hline &\\
			$\lambda_{(1,2)}$ & $\frac{3}{4q}\left(1 + q (w_m -1) + w_m \mp \frac{1}{\sqrt{3(3 + 2 \omega_{\rm BD})}} \,\Delta\right)$ \\[10pt]
			\hline &\\
			$w_{\rm eff}$ & $\frac{w_m(q-1) -1}{q} = \frac{n(w_m -1) + w_m + \alpha}{1 - \alpha + 2n}$\\[10pt]
			\hline &\\
			$\Omega_m$ & $1 - \frac{7 + 3w_m}{2 q} - \frac{3(1 + 2\omega_{\rm BD})(1 + w_m)}{2 q^2} = \frac{(1 + \alpha)(2 + n - 3 n w_m - 2 \alpha)}{2 (1 + 2n - \alpha)^2}$ \\[5pt]
			\hline
		\end{tabular}
		\caption{Behavior of the fifth an sixth fixed points.}
		\label{Tab:Value56}
	\end{center}
	
\end{table}
\noindent where, $\Delta$ is given by
\begin{eqnarray}\label{Eq:delta}
\Delta &\equiv& \sqrt{\begin{aligned}
	&16 q^3 (3w_m -1) + q^2 (17 - 210 w_m - 63w_m^2 + 6 \omega_{\rm BD}(w_m -1 )(7 + 9 w_m)) \\
	&+ 6 q (1 + w_m)(-29 - 3w_m + 2 w_0(6 w_m^2 + 19 w_m -17))- 3 (1 + w_m)^2 \\
	&(48\, \omega_{\rm BD}^2\, (w_m -1) + \omega_{\rm BD}(24 w_m - 74 ) - 27).
	\end{aligned}}
\end{eqnarray}

In order to describe the physical state, the energy fraction of the additional matter has to be greater than zero, i.e.,  $\Omega_m > 0$. This implies that

\begin{eqnarray}\label{Eq:Physical56}
	w_m > \frac{1}{q - 1} = \frac{n - \alpha}{n + 1}.
\end{eqnarray}

In order to understand the behavior, consider the case $(n, \alpha, w_m) = (2, -3/2, w_m)$, as before. In this case,

\begin{eqnarray}
	x^*_{(5, 6)} = -\frac{21}{13} (1 + w_m), \quad y^*_{(5, 6)} = \mp \frac{1}{13} \sqrt{108 w_m^2 - 63 w_m  + 11}
\end{eqnarray}
and the corresponding eigenvalues are
\begin{eqnarray}
	\lambda_{(1, 2)} = \frac{1}{26} \left(30 w_m - 9 \mp \sqrt{697 - 4596 w_m + 9972 w_m^2 - 5184 w_m^3}\right).
\end{eqnarray}

Also, the equation of state, as well as the matter energy fraction, are
\begin{eqnarray}\label{Eq:EEFES56}
	w_{\rm eff} = \frac{1}{13}( 6 w_m - 7), \quad \Omega_m = \frac{1}{169}(6 w_m -7) 
\end{eqnarray}
which implies that, in order to make the fixed points  (\ref{Eq:FixedPoint5}), (\ref{Eq:FixedPoint6}) physical, $w_m > 7/6$ which can be verified from (\ref{Eq:Physical56}). Surprisingly, for $w_m = 7/6$, one of the eigenvalues vanishes. If $w_m$ crosses the value of $7/6$, i.e., $w_m > 7/6$, then, not only the points become physical but also both the eigenvalues become positive. This again implies that the contracting point (\ref{Eq:FixedPoint5}) is stable. This result is {\it remarkable} as when $w_m < 7/6$, as one can remember, the first point (\ref{Eq:FixedPoint1}) becomes an attractor for this case. This point is governed only by the scalar field. However, at this time, the fifth (\ref{Eq:FixedPoint5}) and the sixth (\ref{Eq:FixedPoint6}) points behave as non-physical. At $w_m = 7/6$, (\ref{Eq:FixedPoint1}) becomes saddle as one of the eigenvalues becomes zero. When $w_m$ crosses the values and becomes $> 7/6$, the fifth (\ref{Eq:FixedPoint5}) and the sixth (\ref{Eq:FixedPoint6}) points become physical and the eigenvalues becomes positive, leading to the fifth point (\ref{Eq:FixedPoint5}) to be a stable point. 

The seventh fixed point (\ref{Eq:FixedPoint7}) again is the solution for $V_0 = 0$, i.e., zero potential solution since $y^*_7 = 0$ with the presence of additional matter. The corresponding eigenvalues, effective equation of state and the energy fraction of the additional matter are given in table \ref{Tab:Value7}:

\begin{table}[H]
	\begin{center}
		\begin{tabular}{|c|c|}
			\hline
			&\\
			$x_{7}^*$ &  $\frac{2(1 - 3w_m)(n - \alpha)^2}{n^2 (3w_m -1) + 2n\alpha ( 1 - 3w_m) + \alpha (1 - w_m) + 2 \alpha^2 w_m}$ \\[10pt]
			\hline
			&\\
			$ y^*_{(5, 6)}$ & $0$ \\[10pt]
			\hline &\\
			$\lambda_{1}$ & $\frac{q(1- 3w_m) + 2 \omega_{\rm BD} (1 - w_m^2) + 3 (1 + w_m)}{2 - 2\omega_{\rm BD}(w_m -1)}$ \\[10pt]
			\hline &\\
			$\lambda_{2}$ & $-\frac{2(2- 3w_m) + 3\omega_{\rm BD}(w_m -1)^2}{2 - 2\omega_{\rm BD}(w_m -1)}$ \\[10pt]
			\hline &\\
			$w_{\rm eff}$ & $\frac{\omega_{\rm BD}\,w_m(w_m -1) - 1}{\omega_{\rm BD}\,(w_m -1) - 3}$\\[10pt]
			\hline &\\
			$\Omega_m$ & $1 + \frac{1 - 3w_m}{1 + \omega_{\rm BD}(1 - w_m)} - \frac{\omega_{\rm BD}(1 - 3w_m)^2}{(1 + \omega_{\rm BD}( 1 - w_m))^2}$ \\[5pt]
			\hline
		\end{tabular}
		\caption{Behavior of the seventh points}
		\label{Tab:Value7}
	\end{center}
	
\end{table}

Consider the same case, in the parameter space $(n, \alpha, w_m) = (2, -3/2, w_m)$. The fixed point (\ref{Eq:FixedPoint7}), corresponding eigenvalues, effective equation of state and the energy fraction take the form
\begin{eqnarray}
	&x^*_7  =& \frac{49(1 - 3 w_m)}{72 w_m - 23}, \quad y^*_7 = 0 \\
	&\lambda_1 =& \frac{10 - 69 w_m + 108 w_m^2}{72 w_m - 23}, \quad \lambda_2 = \frac{11 - 63 w_m + 108 w_m^2}{72 w_m - 23}\\
	&w_{\rm eff} =& \frac{49 - 216 w_m + 216 w_m^2}{3 (72 w_m - 23)}\\
	&\Omega_m =& -\frac{10 - 69 w_m + 108 w_m^2}{(72 w_m - 23)^2}
\end{eqnarray}

In order to be a physical state, i.e.,  $\Omega_m > 0$,  the range of $w_m$ lies in $2/9 < w_m < 5/12$. In this limit, it can be shown that the signs of the eigenvalues are opposite to each other. Therefore, the point (\ref{Eq:FixedPoint7}) is unstable.

In a similar way, we can evaluate the fixed points at infinity which can be obtained by choosing a proper transformation from $(x, y)$ to $(u, v)$ \cite{Hrycyna:2013yia}. Since the motivation is in finding and analyzing contracting Universe solution, we do not consider the scenario in this work.

\section{Revisiting matter contraction: An example}\label{Sec:MatterContraction}

In this section, we consider a contracting model which may provide (nearly) scale-invariant spectra. As mentioned earlier, the essential motivation of using these specific model parameters $(n , \alpha, w_m)$ instead of using different model parameters (e.g., in \cite{Hrycyna:2013hla} and \cite{Hrycyna:2013yia}, authors use $(\omega_{\rm BD}, w_m)$ as the model parameters) is to identify the model with their respective power spectra. Since, under conformal transformation, curvature and tensor perturbations remain invariant for single scalar field model, the scale factor $a(\eta)$ in the Einstein frame alone determines the scale-dependence of the perturbations, which, in our case is parametrized by $\alpha$.

In the previous section, we explored the behavior of the system at the fixed points. In this section, we consider contracting model, especially matter contraction model which leads to nearly scale-invariant spectra and show how the scale factor and other physical quantities behave in the vicinity of the first fixed point (\ref{Eq:FixedPoint1}), i.e. the desired contracting point. Since the fixed point corresponds to $\Omega_m = 0$, i.e., $\rho_m = 0$, deviation from the fixed point implies the presence of additional matter with finite energy density\footnote{The deviation can also occur in the absence of the additional matter if we perturb the scalar field potential. However, in that case, $\rho_m$ can be thought of as the deviation of the energy density from the actual energy density due to the perturbed potential in equation (\ref{Eq:Pot}).}. Therefore, in this section, we study the behavior of the system dominated by the scalar field in presence of an additional matter. In doing so, we consider two cases: with additional dust matter, $w_m = 0$ and relativistic matter $w_m = 1/3$.

Matter contracting model implies that the effective equation of state is $w_{\rm eff} = 0$ which can be verified from the effective equation of state in table \ref{Tab:Value12first}. Therefore, the corresponding value of $n$ is equal to $2$. Also, in case of power law inflation with the scale factor $a(t) \sim t^p$ in the Einstein frame, scalar and tensor spectra are of the form

\begin{eqnarray}
	\mathcal{P}_{\rm s}(k) &=& A_{\rm s} \,k^{n_{\rm s} -1} \nonumber\\
	\mathcal{P}_{\rm T}(k) &=& A_{\rm T} \,k^{n_{\rm T}},
\end{eqnarray}
where, $n_{\rm s} - 1 = n_{\rm T} = -2/(p -1)$. Scale invariant spectra demands $n_{\rm s}  = 1$ which implies $p \rightarrow \infty$. $\alpha$ is related to $p$ by $$\alpha = \frac{p}{1 - p},$$ i.e., the corresponding value of  $\alpha$ is $-1$: the Einsteinian Universe is de-Siter. However, observation suggests near scale-invariance (red-tilted) with $n_{\rm s} \approx 0.96$ \cite{1994ApJ...420..445F, Ade:2015xua, Akrami:2018odb, Aghanim:2018eyx}. This implies $p \sim 100$ and hence, $\alpha \sim -1.01$. In other words, a slight deviation from $\alpha = -1$ results the required near scale-invariance\footnote{The $k$-dependence of the power spectrum remains the same even across the bounce if the modes, in this region remain well outside the horizon.}. Therefore, using these model parameters, i.e., $(n = 2,\, \alpha = -1.01)$, we evaluate the behavior of the solution in the vicinity of the fixed point (\ref{Eq:FixedPoint1}). The objective is to observe the effect of deviation from the fixed point (\ref{Eq:FixedPoint1}) in the dynamics\footnote{Note that, since the deviation causes due to the additional matter, it may generate entropy perturbation. However, since the energy density $\Omega_m$ and consequently the effect of the additional matter decays down, we can neglect the entropy perturbation and the spectrum is completely governed by the adiabatic perturbation.}.

\subsection{Universe filled with dust matter}

The additional dust matter corresponds to $w_m$ = 0 and hence, one can verify that the fixed point (\ref{Eq:FixedPoint1}) is stable as the eigenvalues are positive (see table \ref{Tab:Value12second}). Using the model parameters, one can evaluate the point as 
\begin{eqnarray}\label{Eq:3.5}
	x_1^* = -3.02, \quad y_1^* = -0.5.
\end{eqnarray}

Therefore, using the model parameters $(2, -1.01, 0)$, one can obtain the solution for the (\ref{Eq:LinearizedEq}) as
\begin{eqnarray}\label{Eq:Exam1}
	\delta x(N) &\approx& \left(455.51\, \Delta x - 910.52\, \Delta y\right) e^{3.01 N} + \left(-454.51\, \Delta x + 910.52\, \Delta y\right) e^{1.51 N} \\
	\label{Eq:Exam2}
	\delta y(N) &\approx& \left(227.38\, \Delta x - 454.51\, \Delta y\right) e^{3.01 N} + \left(-227.38 \,\Delta x + 455.51\, \Delta y\right) e^{1.51 N}
\end{eqnarray}
 where $N = {\rm ln}\left(\frac{a(t)}{a_0}\right)$, $a_0$ is the scale factor at initial time and $\delta x (0) = \Delta x, \delta y (0) = \Delta y$. Also, we can linearize the equation (\ref{Eq:ConstrainedAcc}) around the point (\ref{Eq:FixedPoint1}) and it takes the form
 \begin{eqnarray}\label{Eq:H21}
  2\, \frac{\dot{H}}{H^2} = \frac{{\rm d\,ln}\, H^2}{{\rm d}N} \approx  - 3 + 1354 \,\delta x(N) - 2698.54\, \delta y(N).
 \end{eqnarray}
 
 Using equations (\ref{Eq:Exam1}) and (\ref{Eq:Exam2}), we can solve (\ref{Eq:H21}) as

\begin{eqnarray}
	{\rm ln}\,\left(\frac{H^2}{H_0^2}\right) &=& - 3 N + (148.34\, \Delta x - 302.17\,\Delta y)  + (1052\,\Delta x - 2103.21\,\Delta y)\,e^{3.01\,N} + \nonumber\\
	&&(-1200.34\,\Delta x + 2405.38 \,\Delta y)\,e^{1.51\,N}.\nonumber
	\end{eqnarray}
	
This can further be simplified if we assume $(\Delta x, \Delta y) \ll 1$ and the Hubble factor can then be written as
	\begin{eqnarray}\label{Eq:Dustmatterapproximate}
	\left(\frac{H^2}{H_0^2}\right) \approx \left(1 - \Omega_{m1} - \Omega_{m2} \right)\left(\frac{a(t)}{a_0}\right)^{-3}  +\Omega_{m1}\,\left(\frac{a(t)}{a_0}\right)^{-1.49}+ \Omega_{m2}\,\left(\frac{a(t)}{a_0}\right)^{0.01} 
\end{eqnarray}
where, $\Omega_{m1} = -1200.34\,\Delta x + 2405.38 \,\Delta y, \quad \Omega_{m2} = 1052\,\Delta x - 2103.21\,\Delta y$.

This result is interesting. The above equation resembles Hubble equation around the vicinity of the fixed point (\ref{Eq:FixedPoint1}) and it behaves like the Universe contains three different types of matter. The first term in the right-hand side is the conventional dust matter as $H^2 \propto a^{-3}$. The second term mimics the matter with the equation of state $\approx -1/2$. The last term is nearly constant and can be thought of like dynamical dark energy. 

This equation represents the attractor behavior as well as the effect of the deviation from the fixed point solution (\ref{Eq:FixedPoint1}). With the presence of an additional matter with the equation of state $w_m = 0$, i.e., the dust matter, the initial point deviates from (\ref{Eq:3.5}). However, due to contraction, $a(t)/a_0$ gets smaller, the terms corresponding to the deviation, i.e., the second and third term in the right-hand side of the above expression becomes sub-dominant compared to the first term and $\delta x$ and $\delta y$ become negligible. This effectively means that the solution can eventually be approximated as $H^2 \propto a^{-3}$, i.e., the Universe is matter contracting. 

While the leading order solution is matter contracting and the corresponding solution is the near scale-invariance of the perturbations \cite{Wands:1998yp}, the deviation from this, too, may play an important role in the spectra. Depending on the detailed nature of these deviations,  it may lead to features in the perturbations which in turn may improve the fit to the observations \cite{Sreenath:2014nca}. 

At the fixed point (\ref{Eq:3.5}), $\Omega_{m}$ is exactly equal to zero. However, again due to the presence of additional dust matter, we can also evaluate the energy fraction of the  matter density (\ref{Eq:Constrainedxy}) as
\begin{eqnarray}
	\Omega_m &\approx& \left(-0.001\,\Delta x + 0.002\Delta y\right) \left(\frac{a(t)}{a_0}\right)^{1.51} + \left(-0.5\,\Delta x + 1.0\,\Delta y\right)  \left(\frac{a(t)}{a_0}\right)^{3.01}
\end{eqnarray}

\noindent Again, the attractor behavior can be observed in the above expression as, even initially non-zero $\Omega_{m}$ eventually drops down to negligible value due to the contraction of the Universe.
  
\subsection{Universe filled with radiation}

In a similar way, the behavior of the solution in the vicinity of the fixed point (\ref{Eq:3.5}) can be studied for $w_m = 1/3$. The linearized solution of the equations (\ref{Eq:LinearizedEq}) can be obtained as

\begin{eqnarray}
	\delta x(N) &\approx& \left(4.04\, \Delta x - 6.07\, \Delta y\right) e^{1.51 N} + \left(-3.04\, \Delta x + 6.07\, \Delta y\right) e^{2.01 N} \\
	\delta y(N) &\approx& \left(2.02\, \Delta x - 3.04\, \Delta y\right) e^{1.51 N} + \left(-2.02 \,\Delta x + 4.04\, \Delta y\right) e^{2.01 N}
\end{eqnarray}
and the Hubble equation becomes
\begin{eqnarray}
	\left(\frac{H^2}{H_0^2}\right)\approx \left(1 - \Omega_{m1} - \Omega_{m2}\right)\left(\frac{a(t)}{a_0}\right)^{-3} + \Omega_{m1}\left(\frac{a(t)}{a_0}\right)^{-1.49} + \Omega_{m2}\left(\frac{a(t)}{a_0}\right)^{-0.99}
\end{eqnarray}
where $\Omega_{m1} = 10.67\,\Delta x - 16.08\,\Delta y, \quad \Omega_{m2} = -9.52\,\Delta x + 19.05\,\Delta y$.

This is fascinating as one can see that the Universe contains three types of matter, one is the conventional dust matter, the other two are (nearly) fluids with the equation of state $-1/2$ and $-2/3$ respectively. Similar to the previous case, the second and the third terms in the right-hand side of the above equation appear due to the presence of the additional radiation matter. However, those terms eventually become sub-dominant compared to the first term due to the contraction. Also,  the fractional matter energy density (\ref{Eq:Constrainedxy}) can be written as

\begin{eqnarray}
	\Omega_{m} = (1.52 \, \Delta x - 3.06\,\Delta y)\left(\frac{a(t)}{a_0}\right)^{2.01} + (-2.02 \,\Delta x + 4.07\,\Delta y)\left(\frac{a(t)}{a_0}\right)^{1.51}
\end{eqnarray}

\noindent Again, in this case as well, as you can see, even the initial non-zero $\Omega_{m}$ becomes negligible after a moment of time.

\section{Conclusion}\label{Sec:Conclu}

In this work, we considered a non-minimally coupled gravity theory and provided the general condition for having a contracting Universe. We also showed that the contraction of the Universe can also behave like an attractor and provided an example of matter contraction that leads to near scale-invariant spectra and obtained the solution near the desired solution.

Due to the arbitrariness of the initial conditions, attractor models are always preferred over the non-attractor model as non-attractors need extreme fine-tuning. The remarkable achievement of slow-roll inflationary paradigm is not only its explanation of near scale-invariant spectra and solving of horizon and flatness problem but also lies in the attractor behavior and therefore solving the initial value problem. However, the growing interest in finding alternatives to it possesses a challenge of not only finding a suitable theory which satisfies all observational constraints but also being an attractor --- like inflationary paradigm. The bouncing models clearly are the viable popular alternatives. However, it mainly faces the above mentioned difficulties, i.e., not being attractors. In this work, we explored the possibilities of finding those models being attractors.

Since, in the minimally coupled frame, i.e., the Einstein frame, contracting Universe models are usually not attractors, we concentrated on the next simplest non-minimal coupling scenario --- the Brans-Dicke theory \cite{Brans-Dicke1961}. To simplify further, we considered a potential that leads to power law solution of the scale factor --- the power law potential.

In order to study the phase space analysis, we examined the model in the presence of an additional barotropic matter with the equation of state $w_m$. Instead of using the Brans-Dicke parameter $\omega_{\rm BD}$, power law potential parameter, $q$ and the equation of state of the barotropic fluid $w_m$ as the model parameters, we used the exponent of the scale factor in the Brans-Dicke frame $n$, exponent of the scale factor in Einstein frame $\alpha$ and the equation of state of the barotropic fluid as our model parameters. The reason is simple: $n$ signifies the effective nature of the Universe whereas $\alpha$ helps tracking the nature of the perturbations, i.e., the $k$-dependence of the perturbations. We studied the phase space and found that the system provides seven fixed points. Amongst them, while four points correspond to the solution in the absence of the additional barotropic fluid, the other three are the solution in  presence of an additional fluid. We identified our desired power law contracting solution (\ref{Eq:FixedPoint1}) and studied its general conditions of being an attractor solution. We showed that, even a conventional {\it `non-attractor'} contracting Universe model can act as an attractor in this non-minimally coupled frame. In doing so, we provided examples of matter contracting Universe, i.e., $n = 2$.

In order to understand how the system behaves near the fixed point which corresponds to the desired contracting Universe in the absence of the additional fluid, we considered a model which leads to the exact $k$-dependence of the perturbations required by the observations. This can be set by choosing the value $\alpha$ appropriately and setting $\rho_m = 0$. In order to describe the contracting Universe, we chose the value $n = 2$, i.e., the Universe is matter contracting. As mentioned earlier, the fixed point corresponds to $\rho_m = 0$ and any deviation from the point suggests the presence of an additional matter with $\rho_m \neq 0$. We studied the effects of these deviations from the actual solution with the presence of a barotropic fluid characterized by the equation of state $w_m$. We found that, even if the Universe contains two fields: the scalar field and the barotropic fluid, the system behaves as it contains three fluids. However, because of the attractor nature of the fixed point, amongst the three, energy densities of the two effective fluids die down soon enough, leading to the scalar field dominated solution only. However, depending upon the details of these deviations, it may produce features in the power spectrum which in turn may help to improve the fit to the observation \cite{Sreenath:2014nca}.

For simplicity, in this work, we considered the conformal coupling function to be $\varphi$. However,  the choice of such conformal coupling function is not unique. For instance, one may find the coupling function to be $f(\varphi, \, \partial_\mu \varphi \partial^\mu \varphi)$ and accordingly the matter part of the action can be chosen in such a way that the background and the stability remain invariant. However, in the most general scalar-tensor theory in four dimension, {\emph{viz.}} Horndeski theory \cite{Horndeski1974}, apart from the conformal coupling, there also exists a derivative coupling between the Einstein tensor and the kinetic part of the scalar field. The above analysis cannot simply be extended to such coupling. We are currently investigating the stability under such coupling.

This work can be thought of as a preliminary study from the model building perspective. This can  help to construct a more viable model which fits best with the observation while still maintaining the attractor nature. For example, since now we know that matter contracting Universe can also be an attractor, we can construct a viable matter bounce model in the Brans-Dicke frame which may be free from all other pathologies. In fact, instead of constructing a matter bounce Universe, this can be generalized to construct any type of bouncing models which can be an attractor and at the same time, the model falls within the observation constraints. This work is currently under progress.

\section*{Acknowledgment}
I thank Rathul Nath Raveendran for useful and indispensable discussions.       
I thank L. Sriramkumar and S. Shankaranarayanan for useful and important discussions as well as their valuable comments on this work.  I also  wish to
thank the Indian Institute of Technology Madras, Chennai, India, for support through the Exploratory Research Project PHY/17-18/874/RFER/LSRI.

\providecommand{\href}[2]{#2}\begingroup\raggedright\endgroup

\end{document}